\documentclass[doublecol]{epl2} 
\pdfoutput=1

\usepackage{latexsym,amsmath,amssymb}
\usepackage{dcolumn}
\usepackage{bm}
\usepackage[usenames,dvipsnames]{xcolor}
\usepackage{epstopdf}
\usepackage[a4paper,colorlinks=true,linkcolor=Blue,citecolor=Blue,urlcolor=Blue]{hyperref}
\usepackage{xspace}

\newcommand{\bra}[1]{\langle #1|}
\newcommand{\ket}[1]{|#1\rangle}

\newcommand{\e}{\varepsilon}
\newcommand{\En}{\mathcal{E}}
\newcommand{\s}{\sigma}
\newcommand{\G}{\Gamma}
\newcommand{\up}{\uparrow}
\newcommand{\down}{\downarrow}

\newcommand{\Ham}{\skew{3}{\hat}{\mathcal{H}}}
\newcommand{\op}[2]{\skew{#1}{\hat}{#2}}

\newcommand{\vopS}{\skew{3}{\hat}{\bm{S}}}
\newcommand{\opS}{\skew{3}{\hat}{S}}
\newcommand{\vops}{\skew{3}{\hat}{\bm{s}}}

\newcommand{\vopSt}{\skew{3}{\hat}{\bm{S}}^\text{t}}
\newcommand{\opSt}{\skew{3}{\hat}{S}^\text{t}}
\newcommand{\Szt}{S_z^\text{t}}

\newcommand{\opc}{\skew{1}{\hat}{c}}
\newcommand{\opn}{\skew{1}{\hat}{n}}
\newcommand{\opd}{\skew{3}{\hat}{d}}
\newcommand{\veco}[1]{\hat{\bm{#1}}}

\newcommand{\matW}{\bm{W}}

\newcommand{\prob}{\mathcal{P}}

\newcommand{\Vthr}{V_\text{thr}}
\newcommand{\dEn}{\delta\mathcal{E}}
\newcommand{\tfin}{t_\text{fin}}

\newcommand{\via}{\emph{via}\xspace}
\newcommand{\ie}{\emph{i.e.}\xspace}

\newcommand{\be}{\begin{equation}}
\newcommand{\ee}{\end{equation}}
\newcommand{\beq}{\begin{eqnarray}}
\newcommand{\eeq}{\end{eqnarray}}

\usepackage[normalem]{ulem}

\title{Manipulating spins of magnetic molecules: Hysteretic behavior with respect to bias voltage}
\shorttitle{Manipulating spins of magnetic molecules} 

\author{Anna P\l{}omi\'nska\inst{1} \and Maciej Misiorny\inst{2} \and Ireneusz Weymann\inst{1}}
\shortauthor{A. P\l{}omi\'nska \etal}

\institute{
  \inst{1} Faculty of Physics, Adam Mickiewicz University, Pozna\'{n}, Poland\\
  \inst{2} Department of Microtechnology and Nanoscience MC2,
  Chalmers University of Technology, G\"{o}teborg, Sweden}


\abstract{
Formation of a magnetic hysteresis loop with respect to a bias voltage is investigated
theoretically in a spin-valve device based on a single magnetic molecule.
We consider a device consisting of two ferromagnetic electrodes bridged
by a carbon nanotube, acting as a quantum dot, to which a spin-anisotropic molecule is exchange coupled.
Such a coupling allows for transfer of angular momentum between
the molecule and a spin current flowing through the dot, and thus,
for switching orientation of the molecular spin.
We demonstrate that this current-induced switching process exhibits
a hysteretic behavior with respect to a bias voltage applied to the device.
The analysis is carried out with the use of the real-time diagrammatic
technique in the lowest-order expansion of the tunnel coupling of the dot to electrodes.
The influence of both the intrinsic properties of the spin-valve device
(the spin polarization of electrodes and the coupling strength of the molecule to the dot)
and those of the molecule itself (magnetic anisotropy and spin relaxation)
on the size of the magnetic hysteresis loop is discussed.}

\begin{document}

\maketitle

\section{Introduction}
%
Over the past years, nano-devices comprising individual magnetic molecules have proven to be very prospective for applications in information-storing and -processing technologies~\cite{Bogani2008Mar,Bartolome_book}.
The key properties of such molecules to be utilized there are an energy barrier for spin reversal, arising when a molecule exhibits a large~(\mbox{$S>1/2$}) effective spin subject to uniaxial magnetic anisotropy~\cite{Gatteschi_book}, and long relaxation times up to tens of microseconds~\cite{Ardavan2007Jan,Bahr2008Feb,Takahashi2009Feb,Takahashi2011Aug}.
As a result, the molecule is magnetically bistable, and its spin can be switched in a controlled way between two metastable states~\cite{Mannini2009}. Such a control of molecular magnetic state, which basically corresponds to manipulation of a bit of information, can be realized either by application of an external magnetic field or by means of spin-polarized currents~\cite{Timm2006Jun,Misiorny2009Jun,Fransson2009Jun,Loth2010}.
In the latter case, the coupling between the molecular spin and tunneling electrons is instrumental in enabling transfer of angular momentum to/from the molecule~\cite{Misiorny2009Jun} ---the process underlying the current-induced magnetic switching.
Recently, it has been experimentally demonstrated that an especially promising setup allowing for implementation of such a coupling involves a carbon nanotube (CNT) on the top of which a magnetic molecule is grafted~\cite{Urdampilleta2011Jul,Ganzhorn2013Feb,Urdampilleta2015Apr}.
For instance, Urdampilleta \etal\cite{Urdampilleta2011Jul} have shown that a device with non-magnetic electrodes but with two molecules attached to a CNT can still effectively act as a spin valve. The observed spin-valve effect is conditioned there by a relative orientation of magnetic moments of the molecules, which can be changed by an externally applied magnetic field. Interestingly, the device also displayed a hysteretic behavior with respect to this field.

Motivated by these developments,
in this letter we consider the dynamical aspects of current-induced
spin reversal of a single molecule embedded in a magnetic tunnel junction.
For this purpose, using the real-time diagrammatic technique
in the regime of sequential tunneling of electrons through the device,
we study the time evolution of expectation values of relevant spin operators.
We show that the process of spin reversal strongly depends on the 
bias voltage applied to the system, which results in 
formation of a magnetic hysteresis loop with respect to applied bias.
The properties of this hysteretic behavior are thoroughly
analyzed for various different parameters of both the junction,
as well as the molecule itself.
We note that although the time-dependent transport through magnetic molecules
have been a subject of several studies \cite{Timm2006Jun,Misiorny2009Jun,Fransson2009Jun,Loth2010,
Lu2009May,Hammar2016Aug,Plominska2017Apr,Plominska2018Jan},
the effect of magnetic hysteresis with respect to transport voltage
and its corresponding properties remain rather unexplored.
The aim of this paper is to fill this gap.

\section{Theoretical description}
%
\begin{figure}[t!]
	\onefigure[width=1\columnwidth]{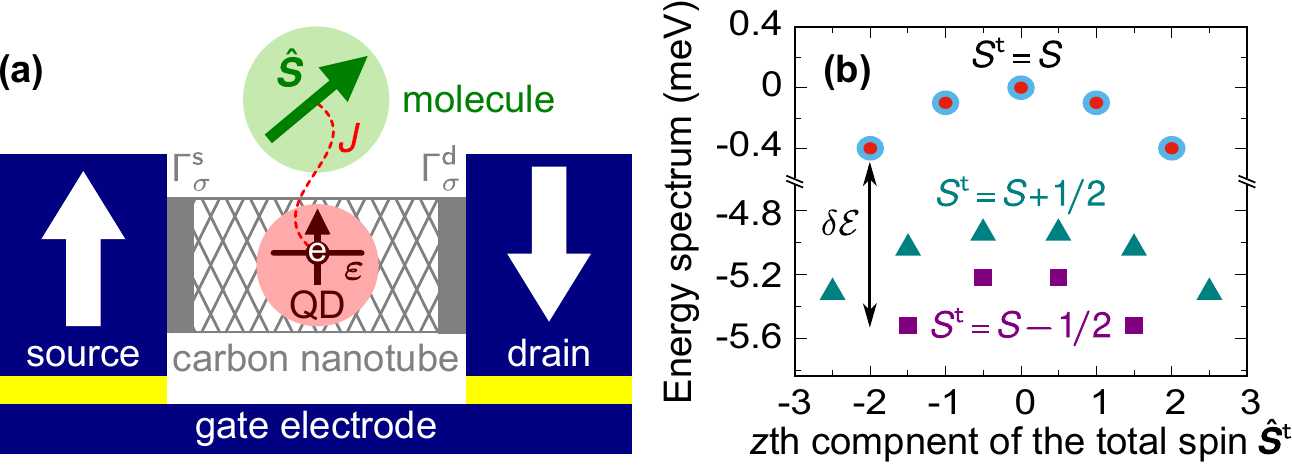}
  	\caption{
  	(Color online)
  	(a) Schematic illustration of a spin-valve device under consideration: A magnetic molecule (represented as a spin~$\vopS$) is grafted on a carbon nanotube, playing a role of a quantum dot (QD), which interconnects two metallic ferromagnetic electrodes.
  	A~gate electrode is used to tune the energy~$\e$ of the QD level.
  	For further details see the main text.
  	(b) Energy spectrum of the QD-molecule system (for~\mbox{$S=2$}) at the particle-hole symmetry point (\mbox{$\e/U=-0.5$}) and for the antiferromagnetic exchange interaction~$J$ between the molecular spin and the electronic spin of the QD (\mbox{$J/U=0.01$}), with \mbox{$U=10$~meV}.
  	The states are labeled by the $z$th component of the total spin~\mbox{$\vopSt=\vopS+\vops$}, with~$\vops$ denoting the QD spin.
    \label{fig1}
  	}
\end{figure}
%

%
The model spin-valve device to be considered consists of a carbon nanotube (CNT) embedded into a magnetic junction, as shown in fig.~\ref{fig1}(a). The CNT operates as a single-level quantum dot (QD) on the top of which a magnetic molecule, represented as an effective large spin~$\vopS$ (\mbox{$S>1/2$}), is deposited.
It is assumed that such a molecule can in general be spin-anisotropic with its spin energetically preferring orientation along some principal axis~($z$), so that an energy barrier for spin reversal arises. Additionally, this spin couples \via exchange interaction to the spin~$\vops$ of electrons tunneling through the QD. As a result, a transfer of angular momentum between the molecule and the tunneling current becomes possible, which essentially constitutes the mechanism of controlling the magnetic state of the molecule~\cite{Misiorny2009Jun}.

%
Formally, the Hamiltonian of the device capturing its key features can be expressed as:
$
	\Ham
	=
	\Ham_\text{jun}
	+
	\Ham_\text{QD}
	+
	\Ham_\text{QD-jun}
	+
	\Ham_\text{mol}
	+
	\Ham_\text{QD-mol}
$.
The first term of~$\Ham$ describes a bare junction formed by two metallic ferromagnetic electrodes (both made of the same material), which are modeled as reservoirs of noninteracting electrons,
$
	\Ham_\text{jun}
	=
	\sum_{qk\s}
	\e_{k\s}^q
	\opc^{q\dag}_{k\s}
	\opc_{k\s}^{q}
$.
The operator $\opc^{q\dag}_{k\s}\ (\opc^{q}_{k\s})$ is responsible for creation (annihilation) of a spin-$\s$ electron with momentum~$k$ and the energy $\e_{k\s}^q$ in the~$q$th electrode, with \mbox{$q=\text{s(ource)},\text{d(rain)}$}.
Moreover, to enable full reversal of the molecular spin, we assume that spin moments of electrodes are antiparallel with respect to each other~\cite{Misiorny2009Jun}, see fig.~\ref{fig1}(a).
The single-level QD is characterized by the next term,
$
	\Ham_\text{QD}
	=
	\e
	\sum_\s
	\opn_\s
	+
	U
	\opn_\up
	\opn_\down
$.
Here, \mbox{$\opn_\s\equiv\opd_\s^\dagger\opd_\s^{}$} stands for the occupation operator counting electrons of spin~$\s$ and energy~$\e$ created (annihilated) in the dot with the operator~$\opd_\s^\dagger\ (\opd_\s^{})$, and~$U$ is the charging energy. Note that~$\e$ can be   adjusted by applying a voltage to a gate electrode.
Finally, tunneling of electrons between electrodes and the QD is given by
$
  \Ham_\text{QD-jun}
  =
  \sum_{qk\sigma}
  \big[
  \sqrt{\G_\s^q/(2\pi\rho_\s^q)}
  \opc^{q\dagger}_{k\sigma}
  \opd_{\sigma}^{}
  +
  \text{H.c.}
  \big]
$,
where the hybridization function~$\G_\s^q$ describes the tunnel coupling between the QD and the $q$th electrode, and $\rho_\s^q$ is the spin-dependent density of states in this electrode. Note also that~$\G_\s^q$ determines the broadening of the QD level.
Introducing the spin-polarization coefficient~$p_q$, defined as
$	
	p_q
	=
	(\G_+^q-\G_-^q)/(\G_+^q+\G_-^q)
$
[with \mbox{$\s=+(-)$} referring to tunneling of spin-majority (-minority) electrons], the hybridization function can be parameterized as \mbox{$\G_\pm^q = (\G_q/2)(1\pm p_q)$}, with $\G_q = \G_+^q+\G_-^q$. Specifically, for the antiparallel magnetic configuration of the junction: \mbox{$\s=+(-)$} corresponds to spin-up (-down) electrons for~\mbox{$q=\text{s}$}, and to spin-down (-up) electrons for~\mbox{$q=\text{d}$}.
Importantly, the system is taken to be fully symmetric with  \mbox{$\G_\text{s}=\G_\text{d}\equiv \G$}, and consequently, \mbox{$p_\text{s}=p_\text{d}\equiv p$}.

%
Furthermore, the magnetic behavior of the molecule grafted on a CNT [see fig.~\ref{fig1}(a)] is included \via the giant-spin Hamiltonian~\cite{Gatteschi_book},
\mbox{$
	\Ham_\text{mol}
	=
	-D\opS_z^2
$},
with \mbox{$D>0$} being the relevant magnetic anisotropy constant.  To keep the discussion simple, we additionally assume that the orientation of the magnetic principal~($z$) axis coincides with that of spin moments in electrodes.
%
%
Last but not least, the exchange interaction between the molecular spin~$\vopS$ and the spin~$\vops$ of electrons tunneling through the QD,
\mbox{$
	\vops
	\equiv
	(1/2)
	\sum_{\s\s^\prime}
	\bm{\sigma}_{\s\s^\prime}^{}
	\opd_\s^\dag
	\opd_{\s^\prime}^{}
$}
with~$\bm{\sigma}$ being the vector of Pauli matrices, has the form
\mbox{$
	\Ham_\text{QD-mol}
	=
	J\veco{S}\cdot\veco{s}
$}.
In this study, the exchange coupling parameter~$J$ is taken to be positive (\mbox{$J>0$}), meaning that the coupling is antiferromagnetic~\cite{Urdampilleta2011Jul}.

\section{Method}
%
In order to analyze the dynamical aspects of hysteretic behavior of the spin-valve device under consideration, we calculate the time evolution of the expectation values~$\langle\Szt\rangle(t)$, $\langle S_z\rangle(t)$ and $\langle s_z\rangle(t)$, corresponding to the~$z$th~component of the total~(\mbox{$\vopSt=\vopS+\vops$}), molecular~($\vopS$) and QD~($\vops$) spin operators, respectively. These can be obtained from
$
	\langle X\rangle(t)
	=
	\sum_\chi
	\bra{\chi}
	\op{2}{X}
	\ket{\chi}\prob_\chi(t)
$
(with \mbox{$\op{2}{X}=\Szt,S_z,s_z$}), where $\prob_\chi(t)$ describes the probability of finding the QD-molecule system at time~$t$ in the eigenstate~$\ket{\chi}$, with \mbox{$\Ham^\prime \ket{\chi} = \En_\chi \ket{\chi}$} and
$
	\Ham^{\prime}
	=
	\Ham_\text{QD}
	+
	\Ham_\text{mol}
	+
	\Ham_\text{QD-mol}
$.
These probabilities $\prob_\chi(t)$ can be found by solving iteratively in time the following master equation,
\begin{equation}
	\bm{\prob}(t+\text{d}t)
	=
	\bm{\prob}(t)
 	+
 	\matW
 	\bm{\prob}(t)
 	\text{d}t
 	,
\end{equation}
for a specified initial condition \mbox{$\bm{\prob}(t=0) = \bm{\prob}_0$}, where the vector $\bm{\prob}(t)$ is formed by probabilities~$\prob_\chi(t)$. The off-diagonal elements of the matrix~$\matW$ are the relevant transition rates, while the diagonal ones account for the probability outflow from a state~$\ket{\chi}$,
$
	W_{\chi\chi}
	=
	\mbox{$-
	\sum_{\chi^\prime(\neq\chi\chi)}
	W_{\chi^\prime\chi}
	$}
$.
In our considerations,~$\matW$ is assumed to be constant in time.
Moreover, $\matW$ is composed of two contributions,
$
	\matW
	=
	\mbox{$\matW^\text{tun}
	+
	\matW^\text{rel}
	$}
$.
The first term, $\matW^\text{tun}$, represents the process of sequential tunneling of electrons between the QD and electrodes, and its off-diagonal elements can be derived with the aid of the real-time diagrammatic technique~\cite{Schoeller1994Dec,Thielmann2005Sep},
\begin{align}
   W_{\chi\chi^\prime}^\text{tun}
   =
   \sum_{q=\text{s,d}}
   \sum_{\s}
   \frac{\G_\s^q}{\hbar}
   \Big\{
   f_q(\En_\chi - \En_{\chi^\prime})
   \big|
   \bra{\chi}
   \opd_{\s}^\dagger
   \ket{\chi^\prime}
   \big|^2
   &
\nonumber\\[-2pt]
   +
   \big[
   1-f_q(\En_{\chi^\prime} - \En_{\chi}) \big]
   \big|
   \bra{\chi}
   \opd_{\s}
   \ket{\chi^\prime}
   \big|^2
   &
   \Big\}
   .
\end{align}
The function
\mbox{$
	f_q(\En)
	\!
	=
	\!
	\big\{
	\!
	\exp[(\En-\mu_q)/(k_\text{B}T)]
	\!
	+
	\!
	1
	\big\}^{\!-1}
$}
stands for the Fermi-Dirac distribution in the $q$th electrode at temperature~$T$ ($k_\text{B}$ ---the Boltzmann constant) and with the electrochemical potential~$\mu_q$.
On the other hand, the second term,~$\matW^\text{rel}$, takes into account the effect of all sources of spin relaxation affecting the dot and the molecule~\cite{Gatteschi_book}. Phenomenologically, such relaxation processes can be described by the effective relaxation time~$\tau$~\cite{Weymann2006May,Misiorny2008May}, and 
\begin{equation}
	W_{\chi\chi^\prime}^\text{rel}
	=
	\frac{\eta_{\chi\chi^\prime}}{\tau}
	\cdot
	\frac{\exp\big[(\En_{\chi^\prime}-\En_\chi)/(2k_\text{B}T)\big]
	}{
	2\cosh\big[(\En_{\chi^\prime}-\En_\chi)/(2k_\text{B}T)\big]
	}
	,
\end{equation}
where
\mbox{$
	\eta_{\chi\chi^\prime}
	\equiv
	\delta_{N(\chi),N(\chi^\prime)}
	\big[
	\delta_{S_z^\text{t}(\chi)-1,S_z^\text{t}(\chi^\prime)}
	+
	\delta_{S_z^\text{t}(\chi)+1,S_z^\text{t}(\chi^\prime)}
	\big]
$}
captures the relevant selection rules, \ie:
(i) relaxation processes do not alter the charge of the QD, defined as
$
	N(\chi)
	=
	\sum_\s
	\bra{\chi}
	\opn_\s
	\ket{\chi}
$;
(ii) the $z$th component of the QD-molecule spin,
$
	\Szt(\chi)
	=
	\bra{\chi}
	\opSt_z
	\ket{\chi}
	,
$
is changed by no more than one quantum of angular momentum due to such processes.

\section{Numerical results and discussion}
%
In the following, we analyze the dynamical aspects of spin-dependent transport through a spin-valve device containing a hypothetical magnetic molecule with spin~\mbox{$S=2$}. The exchange coupling between the QD
and the molecule is assumed to be $J = 0.1$~meV \cite{Urdampilleta2011Jul},
while for the molecule's magnetic anisotropy constant we 
take $D = 0.1$~meV, if not stated otherwise.
All calculations are carried out at temperature \mbox{$k_\text{B}T=0.5$}~meV
and at the particle-hole symmetry point~(\mbox{$\e=-U/2$}),
assuming the charging energy \mbox{$U=10$}~meV \cite{Candini2011Jun}.
Moreover, the coupling between external electrodes and the QD is taken to be \mbox{$\G=0.01$}~meV, and the bias voltage~$V$ is applied symmetrically to the source and drain electrodes~(\ie, \mbox{$\mu_\text{s(d)}=\mp |e|V/2$}).

As already mentioned, due to the exchange coupling (below referred to as the `$J$-coupling') between spins of electrons tunneling through the QD and the molecular spin, the latter can be stabilized in a specific direction along the molecular principal~($z$) axis. Whether the spin of the molecule gets oriented parallel or antiparallel with respect to its principal axis is determined by the polarity of~$V$~\cite{Misiorny2009Jun}, or in other words, by the direction in which the spin-polarized current flows through the device.
Importantly, by changing the polarity of~$V$, the orientation of the molecular spin can be reversed.
However, such a magnetic switching process can be initiated only if the bias voltage~$V$ exceeds 
its threshold value~$\Vthr$, \mbox{$|V|\gtrsim\Vthr$}, which can be related to some activation energy~$\dEn$.
Specifically, in the case under consideration for the antiferromagnetic $J$-coupling (\mbox{$J>0$}), one finds~\mbox{$\Vthr=2\delta\En/|e|$} (with the factor 2 stemming from the symmetric application of a bias voltage to electrodes), and
\begin{equation}
	2\dEn
	=
	U-J/2-D(2S-1)+\Delta
	,
\end{equation}
where
\begin{equation}
	\Delta
	=
	\sqrt{D(D-J)(2S-1)^2+(J/2)^2(2S+1)^2}
	.
\end{equation}
The activation energy $\dEn$ essentially describes here the energy required to change the charge state of the QD-molecule system by one electron. It is indicated in fig.~\ref{fig1}(b) as the energy gap between the lowest-in-energy state of the QD-molecule spin multiplet corresponding to the QD occupied by a single electron (squares) and the lowest-in-energy state associated with the empty/doubly occupied QD (dots/circles).
On the other hand, for~\mbox{$|V|\lesssim \Vthr$},
the rate of processes leading to magnetic switching becomes suppressed,
since these processes can then occur only through thermally-activated events.
Instead, the slow relaxation of the magnetic moment of the molecule is mainly observed,
which arises owing to intrinsic spin relaxation~\cite{Gatteschi_book}
and higher-order spin-flip electron tunneling processes~\cite{Weymann2006May}.
In this work, such relaxation is taken into account \via the effective relaxation time~$\tau$.
In consequence, one expects that sweeping a bias voltage should in principle give
rise to a magnetic hysteresis loop with respect to this voltage.
As we show below, such a hysteresis is a dynamical effect,
and the characteristic time scale at which it can be observed is conditioned by~$\tau$,
as well as by the key parameters of the device, such as,
the $J$-coupling and the magnetic anisotropy constant~$D$.

\subsection{Time evolution of the spin}

%
\begin{figure}[t!]
	\onefigure[width=1\columnwidth]{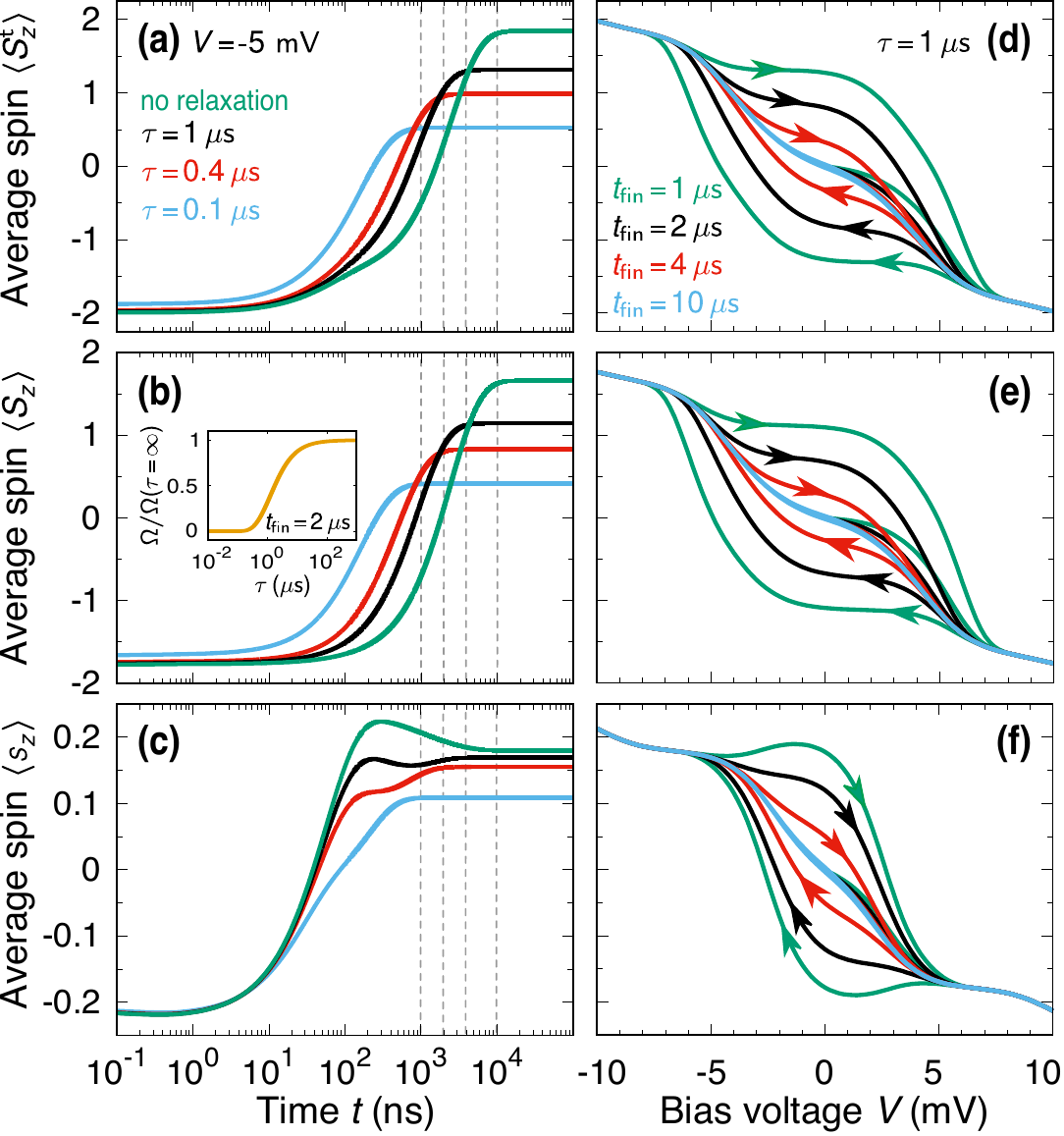}
  	\caption{
  	(Color online)
  	\emph{Left column}:
  	Time evolution of the expectation value~\mbox{$\langle X\rangle\equiv\langle X\rangle(t)$} for the $z$th component of: (a)~the spin of th QD-molecule system (\mbox{$X\equiv\Szt$}), (b)~the~spin of the~molecule (\mbox{$X\equiv S_z$}), and (c)~the~QD spin (\mbox{$X\equiv s_z$}), shown for \mbox{$V=-5$}~mV and selected values of the effective relaxation time~$\tau$ indicated in~(a).
  	\emph{Right column}: Corresponding hysteresis loops for~\mbox{$\tau=1$}~$\mu$s taken at \mbox{$t=\tfin$}, with values of~$\tfin$ given in~(d) and also marked with thin dashed lines in panels (a)-(c).
  	Arrows in panels (d)-(f) indicate the direction of the voltage sweep.
	The inset in~(b) presents the dependence of the area~$\Omega$ of the hysteresis loop for the molecular spin on the relaxation time~$\tau$ (taken at~$\tfin=2$~$\mu$s), scaled to the area in the absence of spin relaxation, \mbox{$\Omega(\tau=\infty)$}.
    The parameters of the system are: $J=0.1$~meV, $D=0.1$~meV and $p=0.5$ (with remaining ones specified in the main text).
    \label{fig2}
  	}
\end{figure}

To understand the mechanism of spin switching and the formation
of hysteresis loop as a function of bias voltage,
it is important to realize how the spin-dependent tunneling processes
affect the magnetic state of the molecule.
For positive bias voltage (\mbox{$V>0$}), i.e., when the electrons tunnel from the 
drain to source electrode, the rate for spin-down electrons to jump
from the drain to the molecule is much faster than that for spin-up electrons.
On the other hand, the spin-up electrons can much more quickly leave
the molecule to the source electrode compared to the spin-down electrons.
These fast tunneling channels are simply associated with the majority
spin subbands of the ferromagnets.
In consequence of an imbalance between the spin-up and spin-down
electron tunneling processes, a nonequilibrium spin accumulation develops in the QD-molecule system,
so that the  spin of the molecule tends to align with the spin moment of the 
drain electrode, \mbox{$\langle S_z \rangle \to -S$}.
The situation becomes, however, completely reversed when the 
polarization of the bias voltage is opposite.
For \mbox{$V<0$}, there is a fast spin-up (spin-down)
channel for tunneling from the source electrode to the molecule
(from the molecule to the drain electrode), such that 
positive spin accumulation builds up in the molecule, leading to
\mbox{$ \langle S_z \rangle \to S$}.
As we show in the following, the build-up of spin accumulation
is a dynamical effect and, depending on relevant time scales,
it can result in the formation of hysteresis loop
for the molecular spin with respect to the bias voltage.

An exemplary time evolution is presented in the left column of fig.~\ref{fig2},
which shows the $z$th component of the total spin $\langle S_z^\text{t} \rangle$,
magnetic molecule's spin $\langle S_z \rangle$ 
and QD's spin $\langle s_z \rangle$ as a function of time $t$
for different values of relaxation time $\tau$ calculated at \mbox{$V=-5$}~mV.
As an initial state for the time evolution we have assumed the steady-state
distribution\mbox{ $\bm{\prob}_0 = \bm{\prob}(t=\infty)$} taken at \mbox{$V = 10$}~mV.
Let us consider first the case in the absence of spin relaxation \mbox{$\tau =\infty$}.
At small times the total spin is stabilized at 
\mbox{$ \langle S_z^\text{t} \rangle \approx -S$}, as expected for positive bias voltage,
however, as the time elapses, the torque transferred to the QD-molecule
system by spin-polarized current leads to the reversion of the total spin.
This happens at the time scale of the order of a few~$\mu$s,
so that for \mbox{$t\gtrsim 1$}~$\mu$s, \mbox{$ \langle S_z^\text{t} \rangle \to S$},
see fig.~\ref{fig2}(a). A~similar spin-switching
can be clearly seen in the $z$th component of the molecule's
spin and the QD's spin. Note, however, that the time
scale for spin reversal of QD is shorter
compared to that of the molecule.
This effect stems from two facts: (i) the molecular
spin is much larger and, thus, more angular momentum needs to be transferred to rotate it;
 (ii) the molecule is only indirectly coupled to the electrodes and angular-momentum (spin) transfer
occurs through the QD ---the~molecule can rotate its spin only after the dot's spin has been reversed.
The reversal times can strongly depend on the spin relaxation
in the system. On can see that for shorter~$\tau$,
the~maximum achievable average value of the spin in the long time limit becomes reduced and,
consequently, the spin reversal is not complete.
Moreover, while decreasing~$\tau$ clearly lowers the time scale for
reversing the spin of the molecule, it hardly affects 
the time scale for changing the spin of the dot.
Finally, in the limit of very fast relaxation (\mbox{$\tau\to 0$}),
the spin immediately relaxes, \ie, \mbox{$\langle S_z^\text{t} \rangle \approx 0$}.

In the remaining discussion, we assume an experimen\-tally relevant value of the relaxation
time \mbox{$\tau = 1$}~$\mu$s, cor\-re\-sponding to typical~$\tau$ for magnetic molecules
at \mbox{$k_\text{B} T=0.5$}~meV \cite{Schlegel2008Oct}.
Also worthy of note here  is that in general the relaxation time strongly depends on both the temperature
and the strength of coupling to external reservoirs
\cite{Bahr2008Feb,Loubens2008Jul,Rastelli2009Mar,Takahashi2009Feb,Takahashi2011Aug}.

\subsection{Formation of hysteresis loop}

Now, let us consider the mechanism of the formation
of the spin hysteresis loop with respect to the bias voltage.
First, for \mbox{$V=0$}, as an initial distribution of occupation probabilities
for the time evolution,~$\bm{\prob}_0$, we take the Boltzmann distribution \mbox{$\bm{\prob}_0 = \bm{\prob}_\text{eq}$}.
Next, we increase the bias voltage, \mbox{$V>0$}, and analyze how the 
expectation values of the corresponding spins, \ie, $\langle S_z^\text{t} \rangle(\tfin)$,
$\langle S_z \rangle(\tfin)$ and $\langle s_z \rangle(\tfin)$, is contingent on the bias voltage~$V$ and 
the final time of evolution \mbox{$t=\tfin$}, see the right column of fig.~\ref{fig2}.
Because for large enough bias voltages the system reaches the steady-state
for assumed final times (\ie, the spin does not depend on final time any more),
we stop increasing the voltage at \mbox{$V=10$}~mV, and then begin the backward sweep,
gradually decreasing the bias voltage from \mbox{$V=10$}~mV to \mbox{$V=-10$}~mV.
The direction of the voltage change is indicated in the figure by relevant arrows.
For this backward sweep, we now assume that the initial state of the system
is described by \mbox{$\bm{\prob}_0 = \bm{\prob}(t=\infty)$} at \mbox{$V=10$}~mV.
Experimentally, it would correspond to the situation
when one applies a large positive bias voltage to initialize the spin state of the system,
and then, by changing the applied voltage, studies its evolution after time \mbox{$t=\tfin$}.
One can see that with lowering the voltage the expectation values of the corresponding spins
follow the lower branch of the loop,  and only when a large negative bias
is applied  the spin can be reversed. As soon as the full spin reversal takes place,
we again start augmenting the bias voltage
and perform a full forward sweep, changing $V$ from \mbox{$V=-10$}~mV to \mbox{$V=10$}~mV.
As an initial probability distribution for this sweep
we take \mbox{$\bm{\prob}_0 = \bm{\prob}(t=\infty)$} at \mbox{$V=-10$}~mV.
Now, one follows the upper branch of the loop
obtaining the switching only after the bias voltage has 
changed its polarity (sign) and exceeded a certain threshold value.
As a result, by sweeping the voltage back and forth
with appropriate initialization of the system,
one obtains a pronounced hysteresis loop of the total spin,
which is presented in fig. \ref{fig2}(d).
Noticeably, a similar hysteretic behavior 
is observed for both the dot's spin and the spin of the bare molecule,
which are displayed in figs. \ref{fig2}(e) and (f).

As can be seen in the right column of fig. \ref{fig2},
the magnitude of the hysteresis loop strongly depends on the final time
of the system's evolution. For assumed parameters,
the loop closes when \mbox{$\tfin \gtrsim 10$}~$\mu$s, and for larger~$\tfin$
one only observes the change of the spin direction in response to reversing the voltage polarity.
For shorter~$\tfin$, however, the hysteresis loop
forms, and the spin behavior depends on the 
direction of the bias voltage sweep.
The observed hysteresis loop is, thus, clearly a dynamical effect
and it becomes larger for shorter final times~$\tfin$.
For the purpose of further analysis, we take the final time 
to be equal to \mbox{$\tfin = 2$}~$\mu$s.
For this choice of $\tfin$ the molecular system does not reach the stationary state yet,
which gives rise to a pronounced magnetic hysteresis loop effect.

Noteworthily,  the magnitude of hysteretic behavior 
is strongly conditioned by the value of the relaxation time~$\tau$.
The inset in fig.~\ref{fig2}(b) demonstrates how the area of the hysteresis loop
\mbox{$\Omega/ \Omega(\tau=\infty)$} of the molecular spin $\langle S_z \rangle$ changes 
as a function of the relaxation time $\tau$,
where \mbox{$\Omega(\tau=\infty)$} is the area of the hysteresis loop calculated 
in the absence of spin relaxation.
One can see that small $\tau$ results in a suppression of the hysteresis loop,
which is due to the fast relaxation of the molecular system to the stationary state.

\subsection{Dependence on intrinsic parameters of the system}

%
\begin{figure}[t!]
	\onefigure[width=1\columnwidth]{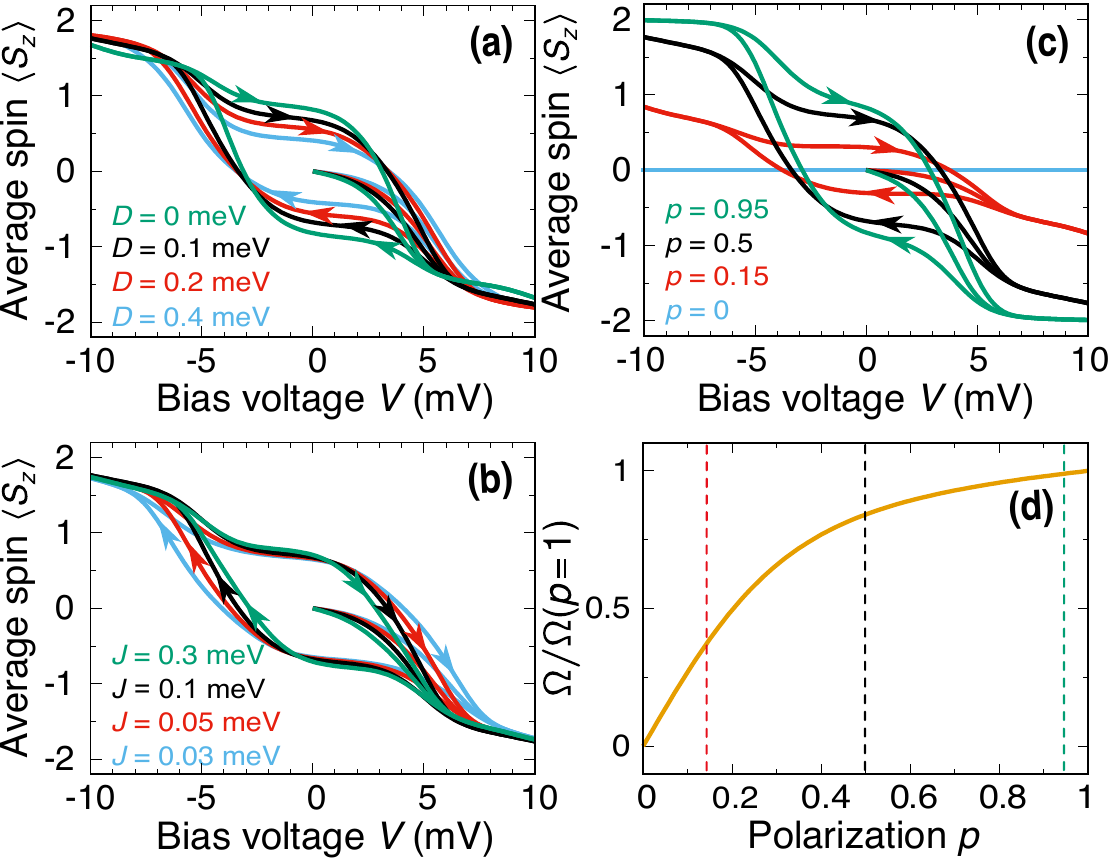}
  	\caption{
  	(Color online)
  	The effect of various internal parameters of the device on the hysteresis loop for the $z$th component of the molecular spin with respect to bias voltage~$V$, calculated at \mbox{$\tfin = 2$}~$\mu$s and \mbox{$\tau = 1$}~$\mu$s. Different loops correspond to selected values of: (a) the magnetic anisotropy constant~$D$, (b) the exchange coupling~$J$, and (c) the spin polarization~$p$ of electrodes.
  	(d) The influence of the spin polarization~$p$ of electrodes on the area~$\Omega$ of  hysteresis loops shown in~(c), normalized to the area obtained for fully polarized electrodes, \mbox{$\Omega(p=1)$}.
	Dashed lines mark the values of $p$ shown in (c).
    Other parameters are the same as in Fig.~\ref{fig2}.
    \label{fig3}
  	}
\end{figure}

We now focus on how the behavior of the hysteresis loop effect changes
when the intrinsic parameters of the device are varied.
The dependence of the expectation value of the molecular spin 
$\langle S_z \rangle$
on the bias voltage for \mbox{$\tau = 1$}~$\mu$s,
calculated for different values of molecule's magnetic anisotropy~$D$,
exchange coupling~$J$ and electrode's spin polarization~$p$,
is shown in fig.~\ref{fig3}.
We begin with the analysis of  the impact of the magnetic anisotropy on the behavior of the hysteresis loop.
It is important to note that $D$ strongly determines the energy spectrum of the molecule.
The increase of $D$ significantly changes the arrangement of the spin multiplets of the molecule,
and gradually leads to their overlap. Moreover, it also
results in an increase of the energy barrier for the process of magnetic switching by spin transfer.
Consequently, the stabilization of the system takes much more time
for larger $D$, and it manifests itself as a gradual increase of 
the bias voltage where the hysteresis loop develops, see
fig.~\ref{fig3}(a).
However, at the same time, the height of the loop
becomes reduced, so that the area of the loop very weakly depends on~$D$
for considered parameters.


The size of the hysteresis loop is also strongly determined
by the strength of the antiferromagnetic \mbox{$J$-inter}\-action
between $\veco{S}$ and $\veco{s}$, see fig.~\ref{fig3}(b).
When the spin-polarized current flows through the QD,
due to this finite exchange coupling, angular momentum can be transferred to the magnetic molecule,
exerting a spin-transfer torque acting on the molecule.
This torque can, in turn, lead to the reversal of the spin of the molecule.
The increase of the strength of the $J$-coupling
facilitates the transfer of angular momentum (spin) to the molecule,
and it accelerates the process of magnetic switching.
A consequence of this effect is the narrowing
of the hysteresis loop as the $J$-coupling becomes larger,
see fig.~\ref{fig3}(b).


The last aspect that we study is the dependence of the hysteresis loop
on the spin polarization $p$ of external leads, which is presented in fig.~\ref{fig3}(c).
Now, one can clearly see the development of hysteretic behavior
as $p$ increases. In the case of non-magnetic leads (\mbox{$p=0$}),
the molecule remains unpolarized, with \mbox{$\langle S_z \rangle = 0$}.
However, already relatively low spin polarization of the electrodes
results in unequal occupation of the molecular spin states,
so that \mbox{$\langle S_z \rangle \neq 0$} and the effect of hysteresis loop
as a function of voltage emerges.
This effect is further enhanced for larger $p$,
and one can observe a nearly perfect switching 
when the leads are close to half-metallic.
The dependence of the magnitude of the hysteresis loop 
on the spin polarization of electrodes is explicitly presented
in fig. \ref{fig3}(d). This figure
shows the area of the hysteresis loop of the magnetic molecule \mbox{$\Omega / \Omega(p=1)$}
as a function of $p$, where \mbox{$\Omega(p=1)$} denotes
the area of the loop for \mbox{$p=1$}.
It can be seen  that the size of the hysteresis grows
relatively fast for small spin polarization,
such as, \mbox{$p\approx 0.2$} with  \mbox{$\Omega/\Omega(p=1) \approx 0.5 $},
while it slows down above \mbox{$p\approx 0.4$},
where one already finds \mbox{$\Omega/\Omega(p=1) \approx 0.8 $},
see fig. \ref{fig3}(d).

\section{Conclusions}

In this letter we have investigated the dynamical aspects of
transport through a molecular spin-valve device
consisting of a CNT-based  QD with an attached molecular magnet~\cite{Urdampilleta2011Jul},
embedded in a magnetic tunnel junction.
The calculations were performed using the real-time diagrammatic technique
in the lowest-order of perturbation expansion with respect to the coupling strength to external leads.
We assumed that the magnetizations of the source and drain
ferromagnetic electrodes form the antiparallel magnetic configuration,
due to  which an imbalance of spin-resolved tunneling processes occurs.
This imbalance results in a spin-transfer torque that can be transferred
to the magnetic molecule and enable the manipulation of its spin.
Consequently, depending on the initial state of the molecule,
the current flow can either result in the stabilization of the molecular spin
or cause its rotation. Changing the direction of the current
flow, it is thus possible to address the spin of the molecule in a desired manner.
Here, we have in particular demonstrated that the spin of the molecule exhibits a hysteretic behavior
with respect to the bias voltage,
which is related to the fact that the process of magnetic switching strongly depends
on the direction of the current flowing through the system.
We have analyzed how the hysteresis loop is affected by various parameters
of the system. First of all, the hysteresis is a dynamical effect,
which disappears for times longer than tens of $\mu$s, and it can be also
suppressed by fast relaxation processes.
Moreover, it turned out that the magnitude of the hysteresis loop 
is greatly conditioned by the spin polarization of the external leads,
the strength of coupling between the QD and the molecule,
but only rather weakly by the value of molecular magnetic anisotropy constant.


\acknowledgments
This work was supported by the National Science Center
in Poland as the Project No. DEC-2013/10/E/ST3/00213.
M.M. also acknowledges financial support from the Polish Ministry of Science
and Education through a young scientist fellowship (0066/E- 336/9/2014) and from the Knut and Alice Wallenberg Foundation.



\begin{thebibliography}{10}
\expandafter\ifx\csname url\endcsname\relax\def\url#1{\texttt{#1}}\fi

\bibitem{Bogani2008Mar}
\Name{Bogani L. \and Wernsdorfer W.} \REVIEW{Nat. Mater.}{7}{2008}{179}.

\bibitem{Bartolome_book}
\Name{Bartolom\'{e} S.~J., Luis F. \and Fern\'{a}ndez J.~F.} (Editors)
  \Book{{Molecular Magnets: Physics and Applications}} NanoScience and
  Technology (Springer, Heidelberg) 2014.

\bibitem{Gatteschi_book}
\Name{Gatteschi D., Sessoli R. \and Villain J.} \Book{{Molecular Nanomagnets}}
  (Oxford University Press) 2006.

\bibitem{Ardavan2007Jan}
\Name{Ardavan A., Rival O., Morton J. J.~L., Blundell S.~J., Tyryshkin A.~M.,
  Timco G.~A. \and Winpenny R. E.~P.} \REVIEW{Phys. Rev.
  Lett.}{98}{2007}{057201}.

\bibitem{Bahr2008Feb}
\Name{Bahr S., Petukhov K., Mosser V. \and Wernsdorfer W.} \REVIEW{Phys. Rev.
  B}{77}{2008}{064404}.

\bibitem{Takahashi2009Feb}
\Name{Takahashi S., van Tol J., Beedle C.~C., Hendrickson D.~N., Brunel L.-C.
  \and Sherwin M.~S.} \REVIEW{Phys. Rev. Lett.}{102}{2009}{087603}.

\bibitem{Takahashi2011Aug}
\Name{Takahashi S., Tupitsyn I.~S., van Tol J., Beedle C.~C., Hendrickson D.~N.
  \and Stamp P. C.~E.} \REVIEW{Nature}{476}{2011}{76}.

\bibitem{Mannini2009}
\Name{Mannini M., Pineider F., Sainctavit P., Danieli C., Otero E.,
  Sciancalepore C., Talarico A.~M., Arrio M., Cornia A., Gatteschi D. \and
  Sessoli R.} \REVIEW{Nat. Mater.}{8}{2009}{194}.

\bibitem{Timm2006Jun}
\Name{Timm C. \and Elste F.} \REVIEW{Phys. Rev. B}{73}{2006}{235304}.

\bibitem{Misiorny2009Jun}
\Name{Misiorny M., Weymann I. \and Barna{\ifmmode \acute{s} \else \'{s}\fi} J.}
  \REVIEW{Phys. Rev. B}{79}{2009}{224420}.

\bibitem{Fransson2009Jun}
\Name{Fransson J.} \REVIEW{Nano Lett.}{9}{2009}{2414}.

\bibitem{Loth2010}
\Name{Loth S., von Bergmann K., Ternes M., Otte A.~F., Lutz C.~P. \and Heinrich
  A.~J.} \REVIEW{Nat. Phys.}{6}{2010}{340}.

\bibitem{Urdampilleta2011Jul}
\Name{Urdampilleta M., Klyatskaya S., Cleuziou J.-P., Ruben M. \and Wernsdorfer
  W.} \REVIEW{Nat. Mater.}{10}{2011}{502}.

\bibitem{Ganzhorn2013Feb}
\Name{Ganzhorn M., Klyatskaya S., Ruben M. \and Wernsdorfer W.} \REVIEW{Nat.
  Nanotechnol.}{8}{2013}{nnano.2012.258}.

\bibitem{Urdampilleta2015Apr}
\Name{Urdampilleta M., Klayatskaya S., Ruben M. \and Wernsdorfer W.}
  \REVIEW{ACS Nano}{9}{2015}{4458}.

\bibitem{Lu2009May}
\Name{Lu H.-Z., Zhou B. \and Shen S.-Q.} \REVIEW{Phys. Rev.
  B}{79}{2009}{174419}.

\bibitem{Hammar2016Aug}
\Name{Hammar H. \and Fransson J.} \REVIEW{Phys. Rev. B}{94}{2016}{054311}.

\bibitem{Plominska2017Apr}
\Name{P{\l}omi{\ifmmode \acute{n} \else \'{n}\fi}ska A., Misiorny M. \and
  Weymann I.} \REVIEW{Phys. Rev. B}{95}{2017}{155446}.

\bibitem{Plominska2018Jan}
\Name{P{\l}omi{\ifmmode \acute{n} \else \'{n}\fi}ska A., Weymann I. \and
  Misiorny M.} \REVIEW{Phys. Rev. B}{97}{2018}{035415}.

\bibitem{Schoeller1994Dec}
\Name{Schoeller H. \and Sch{\ifmmode \ddot{o} \else \"{o}\fi}n G.}
  \REVIEW{Phys. Rev. B}{50}{1994}{18436}.

\bibitem{Thielmann2005Sep}
\Name{Thielmann A., Hettler M.~H., K{\ifmmode \ddot{o} \else \"{o}\fi}nig J.
  \and Sch{\ifmmode \ddot{o} \else \"{o}\fi}n G.} \REVIEW{Phys. Rev.
  Lett.}{95}{2005}{146806}.

\bibitem{Weymann2006May}
\Name{Weymann I. \and Barna\'{s} J.} \REVIEW{Phys. Rev. B}{73}{2006}{205309}.

\bibitem{Misiorny2008May}
\Name{Misiorny M. \and Barna{\ifmmode \acute{s} \else \'{s}\fi} J.}
  \REVIEW{Phys. Rev. B}{77}{2008}{172414}.

\bibitem{Candini2011Jun}
\Name{Candini A., Klyatskaya S., Ruben M., Wernsdorfer W. \and Affronte M.}
  \REVIEW{Nano Lett.}{11}{2011}{2634}.

\bibitem{Schlegel2008Oct}
\Name{Schlegel C., van Slageren J., Manoli M., Brechin E.~K. \and Dressel M.}
  \REVIEW{Phys. Rev. Lett.}{101}{2008}{147203}.

\bibitem{Loubens2008Jul}
\Name{de~Loubens G., Garanin D.~A., Beedle C.~C., Hendrickson D.~N. \and Kent
  A.~D.} \REVIEW{EPL}{83}{2008}{37006}.

\bibitem{Rastelli2009Mar}
\Name{Rastelli E. \and Tassi A.} \REVIEW{Phys. Rev. B}{79}{2009}{104415}.

\end{thebibliography}

\end{document}